**Perspective – Leveraging the Versatility of DNA Origami and Electrochemistry for New Sensing Modalities**

Philip S Lukeman, Department of Chemistry, St. John's University, Queens, NY 11372 USA. lukemanp@stjohns.edu

Abstract: Electrochemical Biosensors are uniquely positioned to offer real-time in vivo molecular sensing due to their robustness to both biofluids and contaminants found in biofluids, and their adaptability for the detection of different analytes by their use of oligonucleotides or proteins as binding moiety. DNA Origami, the folding of a long DNA scaffold by hundreds of shorter oligonucleotide "staple" strands, allows the construction of nanoscale molecular objects of essentially arbitrary form, flexibility and functionality. We describe work at the intersection of these two fields and their – hopefully – bright future together.

**INTRODUCTION**

Noncovalent molecular recognition between two molecules is the primary mechanism by which biosensing is enabled. For a molecular component of a biosensor to work in this manner – as a 'receptor', it must contain a binding site for an analyte, and enable a signal transduction mechanism – a concentration-dependent readout upon binding. Furthermore, for use beyond proof-of-principle, the sensor must readout at relevant concentrations, and the architecture in which it is embedded must be of the appropriate size, and compatible with the physico-chemical environment in which it is embedded. It also must transduce its readout in a timely manner, and must be resistant to spurious signals either from nonspecific binding or other sources of noise from the sample and environment.

A platform which meets these criteria for real-time, *in vivo* biosensing is the use of aptamers, oligonucleotides and antibodies as receptors to construct Electrochemical Biosensors (EBs)[1] (Fig 1A). These platforms consist of a biomolecule i) capable of binding an analyte, ii) modified with redox reporter(s) and iii) attached to a (usually gold) electrode surface. They act as sensors by transducing binding events that result in conformational change - and these changes' resultant effect on redox reporter electron transfer kinetics - into an electrochemical signal. These platforms display many strengths: they are quite robust to biofluids, there are few electrochemically interfering contaminants in vivo, the systems are, in principle, modular - simply swap out one binding site for another in the platform by using a different oligo/protein, there are techniques that reduce drift and the need for external calibration, and the sensors are miniaturizable.

Our expertise is in "structural DNA nanotechnology" [2]; using the self-assembly of polynucleotides via the Watson-Crick base pair to construct objects, lattices and devices. While 10 nm-scale tetrahedral DNA nanostructures, one of the most popular self-assembled DNA objects, have been used in a dizzying array of EBs[3], in this perspective we are going to focus on a much more versatile and profoundly adaptable nanostructure, DNA Origami.

DNA Origami (hereafter referred to as 'Origami')[4], the controlled folding of a long single-stranded "scaffold" DNA by hundreds of designed shorter "staple" oligonucleotide strands, enables the construction of essentially any contiguously-connected two or three-dimensional structure with nm resolution (Fig 1B). Origami is one of the most intrinsically addressable, chemically functionalizable, well-defined, 100 nm-scale molecular architectures currently in existence. Origami has been decorated with numerous types of binding site (aptamer, antibody fragment, small molecule), attachment chemistries and functionality capable of signal transduction…..essentially any moiety that can be conjugated to an oligonucleotide can be appended to Origami[5]. The rigidity and dynamics of Origami can be controlled by judicious design; for example, the binding-induced motion of Origami can be programmed to occur along given axes[6]. Finally, Origami can be instantiated by open source, robust, computer-to-synthesizer pipelines[7]; draw a structure in one of these CAD programs, send the resulting sequences to a DNA synthesis company, and a week later, get the oligos that are required for self-assembly of the object. Origami would thus appear to be excellent candidates for receptor components of sensing platforms; this has proved to be so!

**CURRENT STATUS**

Origami receptors have been probed with gel electrophoresis[8], single molecule methods such as optical tweezers/fluorescence microscopy, and atomic force/electron microscopy[9]; these techniques clearly show, respectively, the ensemble thermodynamics, single-molecule kinetics and structural details of a binding/sensing event. However, these techniques have disadvantages as practical sensors; they are intrinsically slow (tens of minutes at best), subject to interference from spurious binding and contaminants, use bulky instrumentation, and are unusable in complex sample matrices. Origami receptors have also been equipped with circular-dichroic, plasmonic and fluorescence reporters[10] for ensemble measurements ; however, these approaches fail in complex matrices such as blood and other unprocessed biofluids.

Origami has been begun to be explored as a receptor component in some electrochemical platforms. An Origami breadboard has been constructed where electroactive enzymes are displayed on one face of the breadboard; the other face is electrode-bound, and the effect of inter-enzyme distance is measured electrochemically[11]. Similar breadboard architecture underlies a platform for miRNA sensing where the receptors are DNA complements to the miRNA[12]; in an inverted version of this experiment, free Origami in solution acts as an amplifier by sandwiching analyte DNA onto a conventional EB electrode surface[13]. An openable and closeable Origami "zipper" can be electrode bound and used to measure pH[14]; an analogous system uses impedance/capacitance to monitor Origami conformation changes driven by hybridization[15] . All of these approaches show promise; however, all utilize significant concentrations of small-molecule, exogenous redox mediators in the electrolyte, and could not be used under conditions – such as in vivo – that are not compatible with their continual presence.

Other approaches that involve the intersection of Origami and electrochemistry include nanoimpact electrochemical detection of solvated Origami[16], capacitance measurements to determine Origami assembly quality[17] and the utilization of Origami nanopores that sense voltage changes as a function of analyte size, charge and dynamics when these analytes pass through a membrane in which the pore is embedded[18]. It should also be noted that Origami dynamics on an electrode surface can be detected without any redox reporters; Origami can be actuated by an electric field when attached to an electrode and the resultant change in dynamics upon analyte binding can be detected by fluorescence[19]. This system has been used to detect a wide variety of small molecules and proteins; although contemporary platform design eschews Origami for simpler, smaller DNA levers[20] and still displays the limitations of fluorescence based detectors.

None of the Origami systems described above display all of the original strengths of the EB platform; that is, binding and signaling are all on the same molecule, requiring no exogenous reagents, and signaling is driven by conformational change of the specific binding event, thus rendering the system resistant to spurious signals from non-specific binding, and the readout is by biofluid-robust electrochemical signal.

A couple of platforms have used the advantages of Origami and the design principles behind EB sensing and fused them directly. We and the Plaxco group adapted[21] a triangular Origami design[22] to display redox reporters, analyte binding sites, and thiols that enabled its attachment to a gold electrode surface (Fig 1C). As a model analyte system that displayed similar inter-recognition site distances to those on the surface of viruses, we utilized cognate DNA triangles that displayed clusters of 0, 1 or in the case of the target, 3 clusters of 5 binding sites where each cluster was separated by ~40 nm. The sensor successfully discriminated between these three analytes, in a manner that was consistent with reduced flexibility upon polyvalent binding of the triangle; this assessment was supported by Molecular Dynamics simulations of the receptor both free, and bound to monovalent and polyvalent analyte.

Since the publication of this work, flexible assemblies made of multiple rigid multilayer Origami have been demonstrated to bind a whole range of viruses[23] – 'Nanogrippers' have demonstrated similar functionality[24]. If these (or related) Origami could be interfaced with electrochemical detection as described above, they should provide discrimination between viral fragments and whole viruses in a direct manner, that neither PCR nor simple antibody (or other affinity-based approaches) are capable of. Why would you want to detect a whole virus? While unamplified Origami/EB systems will never match the sensitivity and selectivity of PCR, there are cases in which nucleic acid amplification systems provides false positives. For example, spinal fluid of post-polio syndrome patients contains detectable levels of nucleic acids but no virus particles[25] and measles viral RNA is detectable years after the infectious period has passed[26]. Environmental RT-PCR detection of Hepatitis C and other viral RNA was demonstrated with non-infective samples[27]; this is useful for tracking of viral disease in fluids such as wastewater[28]. More generally, detection using genomic amplification should be remarkably scalable for viral threats both known and unknown[29]. However, these polymerase-based techniques fail in measuring the presence of intact virus, and thus, *infectivity* of these fluids in real-time. The fastest cell-culture techniques still take 24h or more[30]; indirect methods have been developed to overcome the limits of PCR in terms of viral capsid integrity[31], and some advances in viral detection have been made with traditional EB sensors[32]. However, given its unique recognition modality, we believe Origami/EB devices have contribution to make to address this gap in both speed and accuracy of analysis of whole, infective viruses.

In another example of the fusion of Origami and EB, in work[33] led by the Rothemund lab, a flat "lily-pad" Origami was tethered, by a long flexible duplex, to a gold electrode (Fig 1D). The bottom of the lilypad displayed dozens of redox reporters, and in a manner analogous to a sandwich assay, also displayed a half analyte-binding domain; the remaining half analyte-binding domain was attached to the electrode surface. Upon configuration as a biosensor, exposure to analyte DNA caused the lilypad to close, bringing reporters close to the surface and generating a signal. The sensor is entirely modular - replacement of the binding site requires a couple of new staples, and optimization for analyte geometry requires swapping out a few staples from a standard library. The resultant sensor displayed similar electrochemical response for the proteins streptavidin and PDGF-BB. In addition, the DNA and streptavidin sensors, after brief optimization as described above, also had class-leading signal gain (for an un-amplified electrochemical system) of hundreds to thousands of percent.

This system is, in this author's opinion, one of the best examples of the modularity of Origami, and should open electrochemical sensing to a wide range of targets from one platform. In conventional EB sensors, the aptamer, oligo or antibody often has to be modified to optimize signal gain while maintaining binding competence. As the binding site and the redox reporter are on the same molecule, a whole range of modifications and re-engineering processes - truncation, destabilization, steric blockage, partially complementary sequences, amongst others, are utilized to this end[34]. For development of selective high-gain sensors, this often means multiple rounds of redesign, resynthesis and retesting of expensive modified oligonucleotides/proteins - which design 'wins' by keeping binding competence while maximizing gain -  is often unintuitive and difficult to explain as electron transfer dynamics are intimately tied to the structure of the single stranded sensing unit. The lilypad approach negates this particular conundrum; binding and signaling function are on separate modular *orthogonal* strands.  Sensor optimization is an easily-modeled function of the analyte geometry, and different geometries are synthesizable and screenable from the same staple library. A more extensive library of lily-pad like designs with adaptors for arbitrary ligand binding and adaptors for target size would enable a genuine "plug-and-play" design for biosensors.

We note that this kind of modular Origami biosensor is not limited to the lily-pad Origami design nor to electrochemical readout. A modular multilayer Origami tweezer-based system has recently been described for use in single-molecule fluorescence DNA and protein detection that displays some of the above advantages[35].

**FUTURE NEEDS**

For use as sensors in biofluids, in vivo and in living animals, many of the same challenges that exist for EBs[36] also need to be overcome for Origami/EB sensors, while others are unique to Origami and its interface with EBs.

From the perspective of the Origami itself, while nuclease activity is a minor component of EB degradation[37], Origami provides a larger attack surface for different nucleases; Origami resistance to nucleases can be engineered[38]. EBs do not need to be protected from denaturation; Origami does - and can be[39] . Questions of toxicology and immunogenicity that have been addressed for oligonucleotides are being addressed by the DNA nanoscience community; for example, immunization with wireframe Origami that displays antigens does not generate antibodies to the wireframe structure in a mouse model[40], nor do these Origami wireframe structures exhibit acute toxicity[41].

Addressing the stability and safety concerns described above, one widely-used Origami-modifying strategy in other, non-biosensor, contexts is the use of pendant, large polymers as a steric shield[38]. In the context of biosensing, the size and chemical impact of these groups will likely result in a case of 'the tail wagging the dog'. The strategic, selective incorporation of crosslinked sites (via processes such as TT welding, where appropriately positioned thymine bases are selectively crosslinked by UV light[42]) and unnatural polynucleotides[43] appear to offer the greatest possibility of retaining the properties of Origami so it can act as an unimpeded biosensor, and not a molecular 'hairball'.

For these structure-switching Origami/EB sensors, many biophysical approaches could be taken to both develop both fundamental understanding of the signal transduction mechanism, and develop optimization heuristics. The most obvious way is to rationally use the exquisite positional control of Origami that allowed the dissection of the distance dependence of FRET to sub-Ångstrom levels[44]. An analogous approach should allow similar dissection of electrochemical signaling in structure-switching Origami/EB, perhaps by using ratiometry as an internal control[45] to study electron transfer behavior at interfaces.

One issue for optimization (for selectivity, robustness or signal strength) by screening of many designs/conditions, is the need to manufacture many sensors. For traditional EBs, as well as serial manufacture of macroelectrodes[46], a variety of microelectrode array manufacture methodologies can be utilized[47]. However, in Origami/EB systems whose sensing mechanism relies on Origami interaction with a smooth e-beam or template-stripped gold electrode surface, the roughness of these electrode surfaces precludes these manufacturing approaches. Lithographic generation of electrochemical microarrays[48] at large scale[49] would appear to offer both the ability to make flat-enough surfaces for ensemble measurements and allow high throughput screening of a range of conditions. More speculatively, as lithographic approaches have allowed arbitrary and absolute orientation of Origami on surfaces[50], electric field anisotropy[51]  could be used to enhance sensor response for oriented Origami.

## CONCLUSIONS

This perspective described recent advances in the overlap and integration of Electrochemical Biosensor and DNA Origami fields.  As EB sensors moved from the lab to clinical testing[1a] to nascent commercialization[52], a wide range of technical problems were addressed by scientists, technologists and clinicians. Similarly, to push Origami/EB advances beyond the proof-of-principle studies described above, this will require collaborations between electrochemists, interfacial scientists, DNA Origami designers, synthetic chemists and biologists to realize the exciting potential of these platforms.

## Acknowledgements

P.S.L. is grateful for support from Army Research Office under Award Number W911NF-231-0283.

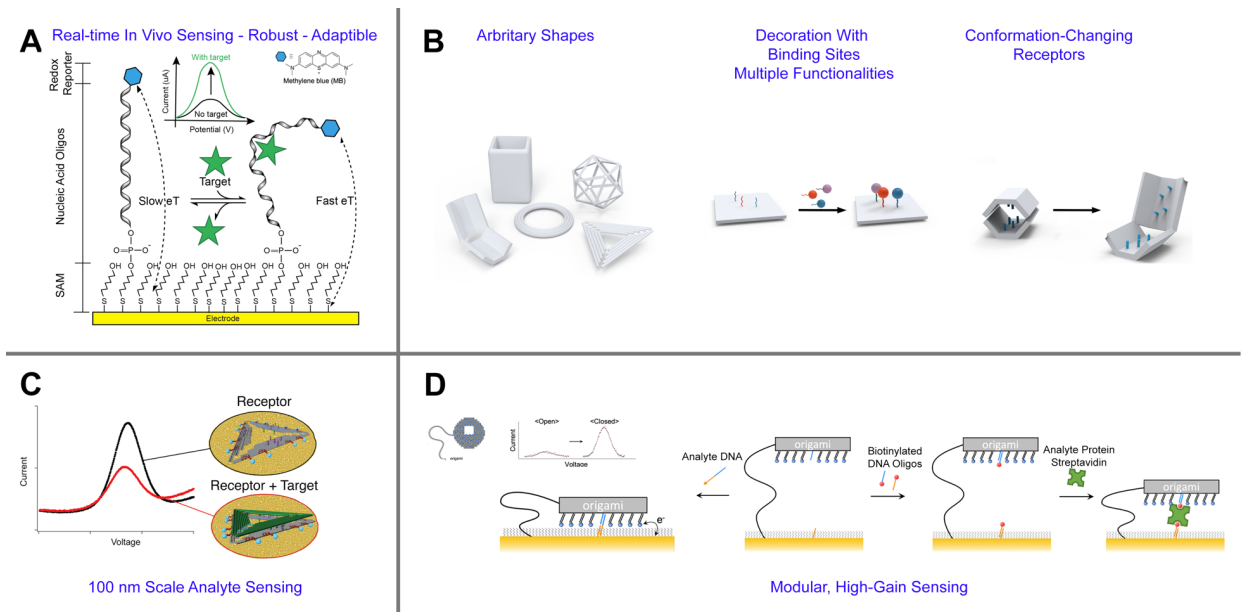

**Figure 1**
**A.** A schematic illustrating the molecular structure and signaling mechanism of electrochemical biosensors (EBs). An EB displays a population of oligonucleotides that undergo electron transfer (eT) at different rates; this rate is affected by target analyte binding, resulting in signal transduction. The inset illustrates the typical target-induced signal change readout of an EB via square wave voltammetry, and the typical reporter used in these systems, Methylene Blue. Adapted from[53] with permission. **B.** DNA Origami's advantages. DNA Origami allows the construction of a vast range of 100 nm-scale objects. These shapes can be functionalized at arbitrary points on their surface with nm resolution, by using functionalized oligonucleotides; these oligos can either be staples or as shown here, bind to staple overhangs. By linking rigid DNA Origami components with flexible single- or double-stranded DNA, devices that undergo large-scale conformational change can be constructed. While renderings are shown in this figure, every structure shown here has been synthesized and verified by atomic force or electron microscopy. Adapted from[54] with permission. **C.** An Origami biosensor capable of recognizing 'mesoscale' objects. Triangular DNA Origami is covalently attached to a gold electrode and functionalized with redox reporters and binding sites. Electrochemical interrogation of this platform successfully monitors binding of 'cognate' 100-nm scale (mesoscale) triangles and differentiates polyvalent from monovalent binding. Binding site spacing on this system is similar in scale to recognition sites on viruses; thus this system acts as a model for recognition of whole-viruses rather than viral fragments. Adapted from[21] with permission. **D.** A modular DNA Origami electrochemical biosensor, adapted to diverse analytes (DNA, protein) by swapping out binding domains. All sensor variants rely on the same 'lilypad' architecture, which undergoes the same 100-nanometer scale conformational change upon binding, bringing dozens of redox reporters close to the electrode, resulting in a sensor that displays class-leading gain. Adapted from[33] with permission.

**References**

[1] a"From the beaker to the body: translational challenges for electrochemical, aptamer-based sensors", N. Arroyo-Currás, *et al.*, *Analytical Methods* **2020**, *12*, 1288-1310; b"Real-time, in vivo molecular monitoring using electrochemical aptamer based sensors: opportunities and challenges", A. M. Downs, *et al.*, *ACS Sensors* **2022**, *7*, 2823-2832; c"An electronic, aptamer-based small-molecule sensor for the rapid, label-free detection of cocaine in adulterated samples and biological fluids", B. R. Baker, *et al.*, *Journal of the American Chemical Society* **2006**, *128*, 3138-3139.

[2] "DNA nanotechnology", N. C. Seeman, *et al.*, *Nature Reviews Materials* **2017**, *3*, 1-23.

[3] "Engineering of Interfaces with Tetrahedra DNA Nanostructures for Biosensing Applications", J. Xuan, *et al.*, *Analysis & Sensing* **2023**, *3*, e202200100.

[4] a"DNA Origami", S. Dey, *et al.*, *Nature Reviews Methods Primers* **2021**, *1*, 1-24; b"Folding DNA to create nanoscale shapes and patterns", P. W. Rothemund, *Nature* **2006**, *440*, 297-302.

[5] "Functionalizing DNA Origami to investigate and interact with biological systems", G. A. Knappe, *et al.*, *Nature Reviews Materials* **2023**, *8*, 123-138.

[6] "DNA Origami nano-mechanics", J. Ji, *et al.*, *Chemical Society Reviews* **2021**, *50*, 11966-11978.

[7] "The art of designing DNA nanostructures with CAD software", M. Glaser, *et al.*, *Molecules* **2021**, *26*, 2287.

[8] "Nanoswitch-linked immunosorbent assay (NLISA) for fast, sensitive, and specific protein detection", C. H. Hansen, *et al.*, *Proceedings of the National Academy of Sciences* **2017**, *114*, 10367-10372.

[9] T. A. Meyer, *et al.*, in *Single-Molecule Tools for Bioanalysis*, Jenny Stanford Publishing, **2022**, pp. 121-172.

[10] "Functionalized DNA Origami-Enabled Detection of Biomarkers", C. Yuan, *et al.*, *ChemBioChem* **2024**, e202400227.

[11] "Constructing submonolayer DNA Origami scaffold on gold electrode for wiring of redox enzymatic cascade pathways", Z. Ge, *et al.*, *ACS applied materials & interfaces* **2018**, *11*, 13881-13887.

[12] "Facile and label-free electrochemical biosensors for microRNA detection based on DNA Origami nanostructures", S. Han, *et al.*, *ACS omega* **2019**, *4*, 11025-11031.

[13] "Signal amplification in electrochemical DNA biosensors using target-capturing DNA Origami tiles", P. Williamson, *et al.*, *ACS Sensors* **2023**, *8*, 1471-1480.

[14] "Probing the conformational states of a pH-sensitive DNA Origami zipper via label-free electrochemical methods", P. Williamson, *et al.*, *Langmuir* **2021**, *37*, 7801-7809.

[15] "Variable Gain DNA Nanostructure Charge Amplifiers for Biosensing", J. M. Majikes, *et al.*, *bioRxiv* **2023**, 2023.2008. 2011.552535.

[16] "Single DNA Origami detection by nanoimpact electrochemistry", E. Pensa, *et al.*, *ChemElectroChem* **2022**, *9*, e202101696.

[17] "Capacitance measurements for assessing DNA Origami nanostructures", V. Walawalkar, *et al.*, *bioRxiv* **2023**, 2023.2003. 2002.530881.

[18] "Solid-state nanopore/channels meet DNA nanotechnology", Q. Ma, *et al.*, *Matter* **2022**.
[19] a"Electrical actuation of a DNA Origami nanolever on an electrode", F. Kroener, *et al.*, *Journal of the American Chemical Society* **2017**, *139*, 16510-16513; b"Magnesium-dependent electrical actuation and stability of DNA Origami rods", F. Kroener, *et al.*, *ACS applied materials & interfaces* **2018**, *11*, 2295-2301.
[20] "Exploring the Interactions of Biologically Active Compounds (Including Drugs) with Biomolecules: Utilizing Surface Plasmon Resonance and SwitchSense Techniques", P. Nowicka, *et al.*, *TrAC Trends in Analytical Chemistry* **2024**, 117764.
[21] "An electrochemical biosensor exploiting binding-induced changes in electron transfer of electrode-attached DNA Origami to detect hundred nanometer-scale targets", N. Arroyo-Currás, *et al.*, *Nanoscale* **2020**, *12*, 13907-13911.
[22] "Immobilization and one-dimensional arrangement of virus capsids with nanoscale precision using DNA Origami", N. Stephanopoulos, *et al.*, *Nano letters* **2010**, *10*, 2714-2720.
[23] "DNA Origami traps for large viruses", A. Monferrer, *et al.*, *Cell Reports Physical Science* **2023**, *4*.
[24] "Designer DNA NanoGripper", L. Zhou, *et al.*, *bioRxiv* **2023**.
[25] "Evidence of presence of poliovirus genomic sequences in cerebrospinal fluid from patients with postpolio syndrome", I. Leparc-Goffart, *et al.*, *Journal of clinical microbiology* **1996**, *34*, 2023-2026.
[26] "Measles virus infections of the central nervous system", U. G. Liebert, *Intervirology* **1997**, *40*, 176-184.
[27] "How stable is the hepatitis C virus (HCV)? Environmental stability of HCV and its susceptibility to chemical biocides", S. Ciesek, *et al.*, *The Journal of infectious diseases* **2010**, *201*, 1859-1866.
[28] "Viruses in wastewater: occurrence, abundance and detection methods", M. V. A. Corpuz, *et al.*, *Science of the Total Environment* **2020**, *745*, 140910.
[29] "A global nucleic acid observatory for biodefense and planetary health", T. N. A. O. Consortium, *arXiv preprint arXiv:2108.02678* **2021**.
[30] "Role of cell culture for virus detection in the age of technology", D. S. Leland, *et al.*, *Clinical microbiology reviews* **2007**, *20*, 49-78.
[31] "Capsid integrity quantitative PCR to determine virus infectivity in environmental and food applications–a systematic review", M. Leifels, *et al.*, *Water Research X* **2021**, *11*, 100080.
[32] "Functional nucleic acid-based biosensors for virus detection", Z. Zhang, *et al.*, *Advanced Agrochem* **2023**, *2*, 246-257.
[33] "Modular DNA Origami-based electrochemical detection of DNA and proteins", B.-j. Jeon, *et al.*, *arXiv preprint arXiv:2312.06554* **2023**.
[34] "Re-engineering aptamers to support reagentless, self-reporting electrochemical sensors", R. J. White, *et al.*, *Analyst* **2010**, *135*, 589-594.
[35] "Engineering Modular and Tunable Single Molecule Sensors by Decoupling Sensing from Signal Output", L. Grabenhorst, *et al.*, *bioRxiv* **2023**, 2023.2011.2006.565795.
[36] "The challenge of long-term stability for nucleic acid-based electrochemical sensors", A. Shaver, *et al.*, *Current opinion in electrochemistry* **2022**, *32*, 100902.

[37] "Nuclease Hydrolysis Does Not Drive the Rapid Signaling Decay of DNA Aptamer-Based Electrochemical Sensors in Biological Fluids", A. Shaver, et al., *Langmuir* **2021**, *37*, 5213-5221.
[38] "Nuclease resistance of DNA nanostructures", A. R. Chandrasekaran, *Nature Reviews Chemistry* **2021**, *5*, 225-239.
[39] "Advancing the Utility of DNA Origami Technique through Enhanced Stability of DNA-Origami-Based Assemblies", S. Manuguri, et al., *Bioconjugate Chemistry* **2023**, *34*, 6-17.
[40] "Enhancing antibody responses by multivalent antigen display on thymus-independent DNA Origami scaffolds", E.-C. Wamhoff, et al., *Nature communications* **2024**, *15*, 795.
[41] "Evaluation of nonmodified wireframe DNA Origami for acute toxicity and biodistribution in mice", E.-C. Wamhoff, et al., *ACS applied bio materials* **2023**, *6*, 1960-1969.
[42] "Sequence-programmable covalent bonding of designed DNA assemblies", T. Gerling, et al., *Science advances* **2018**, *4*, eaau1157.
[43] "Nanostructures from synthetic genetic polymers", A. I. Taylor, et al., *ChemBioChem* **2016**, *17*, 1107-1110.
[44] "Placing molecules with Bohr radius resolution using DNA Origami", J. J. Funke, et al., *Nature nanotechnology* **2016**, *11*, 47-52.
[45] "Dual-reporter drift correction to enhance the performance of electrochemical aptamer-based sensors in whole blood", H. Li, et al., *Journal of the American Chemical Society* **2016**, *138*, 15809-15812.
[46] "High Surface Area Electrodes Generated via Electrochemical Roughening Improve the Signaling of Electrochemical Aptamer-Based Biosensors", N. Arroyo-Currás, et al., *Analytical Chemistry* **2017**, *89*, 12185-12191.
[47] "Electrochemical biosensor arrays for multiple analyte detection", D. Sen, et al., *Analysis & Sensing* **2024**, *4*, e202300047.
[48] "Recent advances in sensor arrays for the simultaneous electrochemical detection of multiple analytes", T. Ozer, et al., *Journal of The Electrochemical Society* **2021**, *168*, 057507.
[49] "Spatially Selective Electrochemical Cleavage of a Polymerase-Nucleotide Conjugate", J. A. Smith, et al., *ACS Synthetic Biology* **2023**, *12*, 1716-1726.
[50] "Absolute and arbitrary orientation of single-molecule shapes", A. Gopinath, et al., *Science* **2021**, *371*, eabd6179.
[51] "Evaluation of electric field in polymeric electrodes geometries for liquid biosensing applications using COMSOL multiphysics", J. A. Gomez-Sanchez, et al., *Sensing and Bio-Sensing Research* **2024**, 100663.
[52] ahttps://diagnosticbiochips.com/; bhttps://www.nutromics.com/.
[53] "Continuous molecular monitoring in the body via nucleic acid–based electrochemical sensors: The need for statistically-powered validation", Y. Yuan, et al., *Current opinion in electrochemistry* **2023**, *39*, 101305.
[54] "Fabricating higher-order functional DNA Origami structures to reveal biological processes at multiple scales", Y. Zhou, et al., *NPG Asia Materials* **2023**, *15*, 25.